\documentclass[pop,superscriptaddress,amsmath,twocolumn]{revtex4}  

\usepackage{graphicx}
\usepackage{gensymb}
\usepackage{dcolumn}
\usepackage{bm}
\usepackage{color}
\usepackage{comment}
\usepackage{amsfonts, amsmath, bm, amssymb,graphicx}

\begin{document}

\newcommand{\beq}{\begin{equation}}
\newcommand{\enq}{\end{equation}}
\newcommand{\half}{\hbox{$1\over2$}}
\newcommand{\bmu}{{\bf\mu}}
\newcommand{\upi}{{\pi}}

\newcommand\Real{\mbox{Re}} 
\newcommand\Imag{\mbox{Im}} 
\newcommand\Rey{\mbox{\textit{Re}}}  
\newcommand\Pran{\mbox{\textit{Pr}}} 
\newcommand\Pen{\mbox{\textit{Pe}}}  
\newcommand\Ai{\mbox{Ai}}            
\newcommand\Bi{\mbox{Bi}}            
\newcommand\E{\mbox{e}}
\newcommand\I{\mbox{i}}
\newcommand\D{\mbox{d}}

\newcommand{\C}{\mathbb C}
\newcommand{\e}{\eqref}
\newcommand{\Lap}[1]{\Delta^{\!\!(#1)}}
\newcommand{\lb}{\left(}
\newcommand{\rb}{\right)}
\newcommand{\lsb}{\left[}
\newcommand{\rsb}{\right]}
\newcommand{\Rin}{R_{i}}
\newcommand{\Xout}{X_{o}}
\newcommand{\Xin}{X_{i}}
\newcommand{\Rout}{R_{o}}
\newcommand{\Lin}[1]{\left\|#1\right\|_{i}}
\newcommand{\Lout}[1]{\left\|#1\right\|_{o}}
\newcommand \mcE{{\mathcal E}}
\newcommand \bfr{{\bf r}}
\newcommand \bfk{{\bf k}}
\newcommand \bfq{{\bf q}}
\newcommand \bfU{{\bf U}}
\newcommand \bfV{{\bf V}}

\def\res{\mathop{\mathrm{Res}}\limits}
\newcommand{\sgn}{\operatorname{sgn}}
\newcommand{\R}{\mathbb R}
\newcommand{\p}{\partial}
\newcommand\Fr{\mbox{\textit{Fr}}}
\newcommand\Ham{\mbox{\textit{H}}}
\newcommand\Arg{\mbox{\textrm{Arg\,}}}
\newcommand\bigO{\mbox{\mathcal{O}}}

\title{The instabilities beyond modulational type in a repulsive Bose-Einstein condensate with a periodic potential}

\author{Wen-Rong Sun}
\email{sunwenrong@ustb.edu.cn}
\author{Jin-Hua Li}
\email{m202210704@xs.ustb.edu.cn}
\affiliation{School of Mathematics and Physics, University of Science and Technology Beijing, Beijing 100083, China}

\author{Lei Liu}
\email{liueli@126.com}
\affiliation{College of Mathematics and Statistics, Chongqing University, Chongqing, 401331, China}
\author{P.G. Kevrekidis}
\email{kevrekid@umass.edu}
\affiliation{Department of Mathematics \& Statistics, University of
Massachusetts, Amherst, MA 01003, USA}

\date{\today}

\begin{abstract}
  The instabilities of the nontrivial phase elliptic solutions in a repulsive Bose-Einstein condensate (BEC)  with a periodic potential are investigated. Based on the defocusing nonlinear Schr\"{o}dinger (NLS) equation with an elliptic function potential,  the well-known modulational instability (MI), the more recently identified
  high-frequency instability, and an unprecedented ---to our knowledge---
  variant of the MI, the so-called isola instability  are identified numerically. Upon varying parameters of the solutions, instability transitions occur through the suitable
  bifurcations, such as the Hamiltonian Hope one. Specifically, (i) increasing
  the elliptic modulus $k$ of the solutions, we find that MI switches to the isola instability and the dominant disturbance has twice the elliptic wave's period, corresponding to a Floquet exponent $\mu=\frac{\pi}{2K(k)}$. The isola instability arises from the collision of spectral elements at the origin of the spectral plane.
  (ii) Upon varying $V_{0}$, the transition between MI and the high-frequency instability occurs. Differently from the MI and isola instability where the collisions of
  eigenvalues happen at the origin, high-frequency instability arises from pairwise collisions of nonzero, imaginary elements of the stability spectrum; (iii) In the limit of sinusoidal potential, we show that MI occurs from a collision of eigenvalues with $\mu=\frac{\pi}{2K(k)}$ at the origin; (iv) we also examine the dynamic byproducts
  of the instability
  in chaotic fields generated by its manifestation.
  An interesting observation is that,  in addition to MI, the isola instability could also lead to dark localized events in the scalar defocusing NLS equation.
\end{abstract}

\keywords{Modulational instability, Bose-Einstein condensates, Optical lattices,
  Water Waves, Stokes Waves}
\maketitle



\section{\label{s:Intro }Background and Motivation}

Bose-Einstein condensates (BECs) trapped in the periodic potentials, such as the one induced by standing light waves (optical lattices) have attracted considerable attention already since the early studies on the subject (summarized, e.g., in~\cite{becbook1,becbook2,morsch}) and even to this day; see, e.g., the recent review of~\cite{bec2-2}. BECs trapped in standing light waves have been applied to investigate such diverse phenomena as phase coherence~\cite{bec3,bec4}, matter-wave diffraction~\cite{bec5}, quantum logic~\cite{bec6,bec7} and so on~\cite{bec8,bec9}.  Since an interplay between periodicity and nonlinearity (even when the interatomic interaction is repulsive), some striking effects appear, such as  localized structures~\cite{bec10,bec10-2,bec10-3} and instabilities~\cite{bec11,bec12,bec13,bec14}.

More specifically, the study of instabilities has been a topic of wide
interest, as illustrated, e.g., in focused reviews on the subject~\cite{instab}.
Indeed, the modulational instability (MI)
has been used experimentally, in conjunction with
a magnetic tuning of condensate interactions (from repulsive to attractive)
as a method for producing bright solitonic trains and observing
their interactions since over 20 years~\cite{hulet1}. The relevant technique
has continued to be at the forefront of experimental developments for a
considerable while with experimental progress revealing more clearly the
nature of solitonic interactions more recently~\cite{hulet2}. Indeed, more
recent studies have enabled a systematic and even quantitative comparison
of experimental outcomes against predictions (e.g., of soliton numbers
created by MI) of effective 1d theoretical/computational models~\cite{robins}.
Another recent dimension of the ever expanding influence and impact
the MI more recently has been its experimental use (in conjunction again
with a quench from repulsive to attractive interactions) in order to produce
---this time in a quasi-two-dimensional setting--- of wavepackets
leading to the famous Townes soliton~\cite{hung}.

On the discrete
(or quasi-discrete) realm of focal interest to this work, the modulational
instability has been central to theoretical and experimental implementations
not only in atomic BECs, but also in other proximal areas of dispersive
wave phenomena. In particular, in BECs in the context of optical lattices the
discrete modulational instability was theoretically proposed~\cite{trombe}
and subsequently experimentally illustrated~\cite{fort} to be responsible
for a dynamical superfluid-insulator transition for an array of weakly coupled
condensates driven by an external harmonic field. Shortly thereafter,
such an instability was reported for the first time in the context of
onlinear optics, using an AlGaAs waveguide array with a self-focusing
Kerr nonlinearity~\cite{christo}. Finally, relevant features have been
leveraged as a means of producing robust nonlinear coherent
structures via MI in other proximal fields featuring discrete media,
such as, for instance, in the case of a diatomic granular crystal
in the work of~\cite{boechler}.

In this paper, we revisit the quasi-discrete setting quasi-one-dimensional repulsive BEC trapped in a periodic potential; see, e.g.,~\cite{bec10-2,bec10-3,bec11,bec12,bec13,bec14,gp1,gp2} for only some
among numerous examples. Our emphasis is on the 
study of instabilities and localized structures numerically, utilizing a numerical
set of tools that have been developed more recently than some of these
important works and which, we believe, reveal a number of unprecedented
features and instabilities in the relevant system, worthwhile of further
---and potentially also experimental, given the recent developments discussed
above--- consideration. We now proceed to formulate the problem
mathematically and discuss some of the main findings.

\section{\label{s:Form} Mathematical Formulation and Main Results}

The governing equation is given by the defocusing NLS model with external potential~\cite{bec13, gp1,gp2}
\begin{equation}
i \psi_t=-\frac{1}{2} \psi_{x x}+|\psi|^2 \psi+V(x) \psi,\label{bec1}
\end{equation}
where $\psi(x,t)$ is the macroscopic wave function of the condensate. Confinement in a standing light wave leads to $V(x)$ being periodic~\cite{bec11,bec12,bec13,bec14}, 
\begin{equation}\label{bec1a}
V(x)=-V_0 \mathrm{sn}^2(x, k),
\end{equation}
where $\mathrm{sn}^2(x, k)$ is the Jacobian elliptic sine function
with elliptic modulus $k\in[0,1]$. When $k=0$, $\mathrm{sn}(x, k)$
becomes $\sin (x)$ and the potential $V(x)$ is a standing light
wave~\cite{morsch}. As discussed in~\cite{bec10-3, bec13}, when
$k<0.9$, the potential $V(x)$ resembles the behavior of $\sin (x)$ and
could provide a good approximation to a standing light wave, while at
the same time retaining the advantage of analytically tractable
solutions
that were leveraged towards a number of analytical results in the
above works.

The stationary condensates are described by the solutions to~(\ref{bec1}) of the form~\cite{bec12,bec13}: 
\begin{equation}\label{solution1}
\psi(x, t)=r(x) \exp (-i \omega t+i \theta(x)),
\end{equation} 
where  $\theta(x)=c \int_0^x \frac{d
  x^{\prime}}{r^2\left(x^{\prime}\right)}$and $r^2(x)=A
\operatorname{sn}^2(x, k)+B$. Besides,  the relations among the parameters $\omega$, $c$, $A$, $B$, $V_0$ and $k$ are $\omega=\frac{1}{2}\left(1+k^2+3 B-\frac{B V_0}{V_0+k^2}\right)$, $c^2  =B\left(1+\frac{B}{V_0+k^2}\right)\left(V_0+k^2+B k^2\right)$ and $A =V_0+k^2$.
To require that $r^2(x)>0$ and $c^2>0$ implies the following conditions:
$V_0 \geq-k^2$ and $B \geq 0$, or
$V_0 \leq-k^2$ and $-\left(V_0+k^2\right) \leq B
\leq-\left(1+\frac{V_0}{k^2}\right)$. We note that $r(x)$ is periodic
with period $2K(k)$. The stability and instability of the trivial
phase elliptic solutions $(c=0)$ have been studied
in~\cite{bec12,bec13,bec14}.  However,
the availability of instability results about the nontrivial phase
elliptic solutions
is far more limited.
This constitutes a fundamental and more concrete motivation
of the present work. Our main corresponding findings are as follows:

(i) It is well known that the modulational instability (MI), also
known
---especially so in the context of fluids--- as the Benjamin-Feir instability, originated from the study of stability of Stokes waves in deep water (in the late 1960s)~\cite{mi1}. Then MI has been predicted and observed in BECs~\cite{hulet1,hulet2,robins,trombe,fort,mi2,mi3,mi4,mi5,mi6} and nonlinear optics~\cite{mi9,mi10,mi11,mi12,mi13,mi14}, as well as in other physical media~\cite{mi15,mi16,mi17}.
In 2011,  Deconinck and Oliveras~\cite{mi18} first displayed the full
stability spectra of Stokes waves in finite and infinite depth. More
importantly, they showed that in addition to MI, another
instabilities, taking place away from the origin of the so-called
spectral plane (the plane of the imaginary vs. the real part of the
corresponding
eigenvalues) exist. The instabilities are also referred to as
high-frequency instabilities.
These high-frequency instabilities have been studied
analytically~\cite{mi19,mi20}.  Therefore, a natural question arises:
Do these high-frequency instabilities (originating in
the fluid setting) exist in BECs? In this paper, we show the existence
of the high-frequency instability in BECs and study the transition
between the high-frequency instability and MI
in the context of the model of Eq.~(\ref{bec1}) with the potential of
Eq.~(\ref{bec1a}).

Additionally, in 2022, the authors of~\cite{ber1} studied the
instability of near-extreme Stokes waves. One important feature
identified
in~\cite{ber1} is the appearance of what we refer to as the isola
instability branch.
For such an instability, the eigenvalues in the case of~\cite{ber1}
correspond to eigenfunctions 
that are localized near the wave crest as the extreme wave
is approached.
Importantly, a telltale sign of such an instability
that we will use to distinguish it from MI is that it
detaches
from the origin of the spectral plane and corresponds to a band of
complex eigenvalues that thereafter remains detached from the origin (contrary to the
case of MI, where the band encompasses the origin).
The transition between isola instability and MI is investigated. 
We expect that such an instability branch may be tractable in BECs,
based on the above significant and quantitative experimental
progress therein~\cite{hulet2,robins,hung}.

(ii) The standard defocusing NLS equation (i.e., $V_{0}=0$
in~(\ref{bec1})) does not admit rogue-wave solutions since all
periodic traveling solutions (including plane-wave solutions) of the
defocusing NLS equation are stable. However, by considering the
external potential in the defocusing regime, i.e.,  $V_{0}\neq0$
in~(\ref{bec1}),  we show that different instabilities
occur. Therefore, a natural question arises:  do spatio-temporal
localization events exist in the defocusing NLS equation with an
elliptic function potential?  It is well known that MI could lead to
the formation of localization events~\cite{1bec,2bec,3bec,4bec,5bec}.
In particular, it is especially interesting (given its recent
identification) to explore whether specifically the {\it isola
  instability} could lead to localized events. Indeed, the present
work
illustrates the dynamical evolutions that showcase how this phenomenon
takes place.

\section{\label{PF} Computational Technique of Choice: Hill's Method
  and its Setup}
The linear stability of~(\ref{solution1}) in the setting of the model
of Eq.~(\ref{bec1}) is explored by considering the following form:
\begin{equation}
\psi(x, t)=(r(x)+\epsilon \phi(x, t)) \exp [i(\theta(x)-\omega t)],
\end{equation}
where $\epsilon \ll 1$ denotes a small parameter. With $\mathbf{U}=\left(U_1, U_2\right)^T=(\operatorname{Re}[\phi], \operatorname{Im}[\phi])^T=\hat{\mathbf{U}}(x) \exp (\lambda t)=\left(\hat{U}_1, \hat{U}_2\right)^T\exp (\lambda t)$,  the eigenvalue problem is expressed as~\cite{bec12,bec13},
\begin{equation}
\mathcal{L} \hat{\mathbf{U}}=\left(\begin{array}{cc}
 -\frac{c}{r(x)} \partial_x \frac{1}{r(x)} &-L_{-}\\
L_{+} & -\frac{c}{r(x)} \partial_x \frac{1}{r(x)}
\end{array}\right) \hat{\mathbf{U}}=\lambda \hat{ \mathbf{U}},\label{stability1}
\end{equation}
where
\begin{align}
L_{+}  =-\frac{1}{2}\left(\partial_x^2-\frac{c^2}{r^4(x)}\right)+3 r^2(x)+V(x)-\omega, \\
L_{-}  =-\frac{1}{2}\left(\partial_x^2-\frac{c^2}{r^4(x)}\right)+r^2(x)+V(x)-\omega,
\end{align}
and $\lambda$ is a complex number, i.e., the corresponding eigenvalue
of the linearization.
We note that when $c=0$, the stability problem~(\ref{stability1})
corresponds to the trivial phase solutions. This case was examined
in~\cite{bec12,bec13}. We only focus on the stability problem of the
nontrivial phase solutions $(c\neq 0)$ using the
so-called Hill's method that was originally theoretically developed
and
computationally implemented in~\cite{ber2}.

Since the the coefficient functions of the stability problem~(\ref{stability1}) are periodic in $x$ with period $L=2K(k)$, we write all coefficient functions as the complex Fourier form, i.e., $r^2(x)=\sum_{n=-\infty}^{\infty} Q_n e^{i 2 n \pi x / L}$, $ r^{-2}(x)=\sum_{n=-\infty}^{\infty} R_n e^{i 2 n \pi x / L}$,
$r^{-4}(x)=\sum_{n=-\infty}^{\infty} S_n e^{i 2 n \pi x / L}$, $r^{-3}(x) r^{\prime}(x)=\sum_{n=-\infty}^{\infty} T_n e^{i 2 n \pi x / L}$,
and $V(x)=\sum_{n=-\infty}^{\infty} \hat{V}_{n} e^{i 2 n \pi x / L}$. Here $Q_{n}$, $R_{n}$, $S_{n}$,  $T_{n}$ and $\hat{V}_{n}$ denote the Fourier coefficients. 
The periodicity of coefficient functions of~(\ref{stability1}) allows
us to decompose the perturbations using Floquet’s Theorem
(see~\cite{ber2} for details)
\begin{align}
\hat{U}_{1}(x)=e^{i \mu x} H_{1}(x)=e^{i \mu x} \sum_{n=-\infty}^{\infty} \hat{U}_{1n} e^{i 2 n \pi x /P L},\\
\hat{U}_{2}(x)=e^{i \mu x} H_{2}(x)=e^{i \mu x} \sum_{n=-\infty}^{\infty} \hat{U}_{2n} e^{i 2 n \pi x / P L},
\end{align}
where the Floquet exponent $\mu\in[0,2\pi/L)$, and
\begin{align}
\hat{U}_{1n}=\frac{1}{P L} \int_{-P L / 2}^{P L / 2} H_{1}(x) e^{-i 2 \pi n x / P L} d x,\\
\hat{U}_{2n}=\frac{1}{P L} \int_{-P L / 2}^{P L / 2} H_{2}(x) e^{-i 2 \pi n x / P L} d x.
\end{align}
Here, we expand $H_{1}(x)$ and $H_{2}(x)$ as a Fourier series in $x$ with period $PL$, where $P\in \mathbb{N}$.
Substituting all of the above Fourier expansions into~(\ref{stability1}) and equating Fourier coefficients lead to the following bi-infinite eigenvalue problem: 
\begin{subequations}\label{st}
\begin{eqnarray}
&&\left(\omega-\frac{1}{2}\left(i \mu+\frac{i 2n \pi}{PL}\right)^2\right) \hat{U}_{2n}- \sum_{m=-\infty}^{\infty} Q_{\frac{n-m}{2}} \hat{U}_{2m}\nonumber\\
&&- \sum_{m=-\infty}^{\infty} \hat{V}_{\frac{n-m}{2}} \hat{U}_{2m}
-\frac{c^2}{2} \sum_{m=-\infty}^{\infty} S_{\frac{n-m}{2}} \hat{U}_{2m} \\
&&+ c \sum_{m=-\infty}^{\infty} T_{\frac{n-m}{2}} \hat{U}_{1m}-c\sum_{m=-\infty}^{\infty} R_{\frac{n-m}{2}} \hat{U}_{1m}=\lambda \hat{U}_{1n},\nonumber\\
&&-\left(\omega-\frac{1}{2}\left(i \mu+\frac{i 2n \pi}{PL}\right)^2\right) \hat{U}_{1n}+3\sum_{m=-\infty}^{\infty} Q_{\frac{n-m}{2}} \hat{U}_{1m}\nonumber\\
&&+ \sum_{m=-\infty}^{\infty} \hat{V}_{\frac{n-m}{2}} \hat{U}_{1m}
+\frac{c^2}{2} \sum_{m=-\infty}^{\infty} S_{\frac{n-m}{2}} \hat{U}_{1m} \\
&&+ c \sum_{m=-\infty}^{\infty} T_{\frac{n-m}{2}} \hat{U}_{2m}-c\sum_{m=-\infty}^{\infty} R_{\frac{n-m}{2}} \hat{U}_{2m}=\lambda \hat{U}_{2n},\nonumber
\end{eqnarray}
\end{subequations}
where $Q_{\frac{n-m}{2}},R_{\frac{n-m}{2}}, S_{\frac{n-m}{2}}, T_{\frac{n-m}{2}}, \hat{V}_{\frac{n-m}{2}} =0$ if $\frac{n-m}{2}\in \mathbb{Z}$.

The bi-infinite eigenvalue problem~(\ref{st}) is equivalent to~(\ref{stability1}).  We will determine the spectrum of the linearized operator about the stationary solutions using the bi-infinite eigenvalue problem~(\ref{st}). The stability spectrum of the elliptic solutions is constructed as the union of the spectra for all values of $\mu$.

\begin{figure}[tbp]
    \centering
    \includegraphics[width=0.25\textwidth]{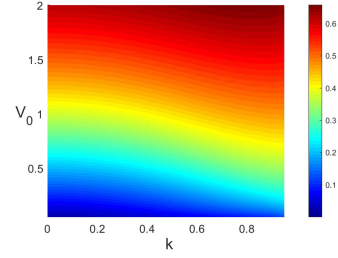}
    \vspace*{-0.1in}
    \caption{The maximal instability growth rate $\gamma$ as a function of $k$ and $V_{0}$ with $B=0.25$.}
    \vspace*{-0.1in}
    \label{fs1}
\end{figure}

\section{\label{IR} Instability Results}
In this section, by choosing a cut-off $N$ on the number of Fourier modes, we numerically find the spectrum to~(\ref{stability1}) using the bi-infinite eigenvalue problem~(\ref{st}).

\begin{figure*}
    \centering
    \includegraphics[width=0.255\textwidth]{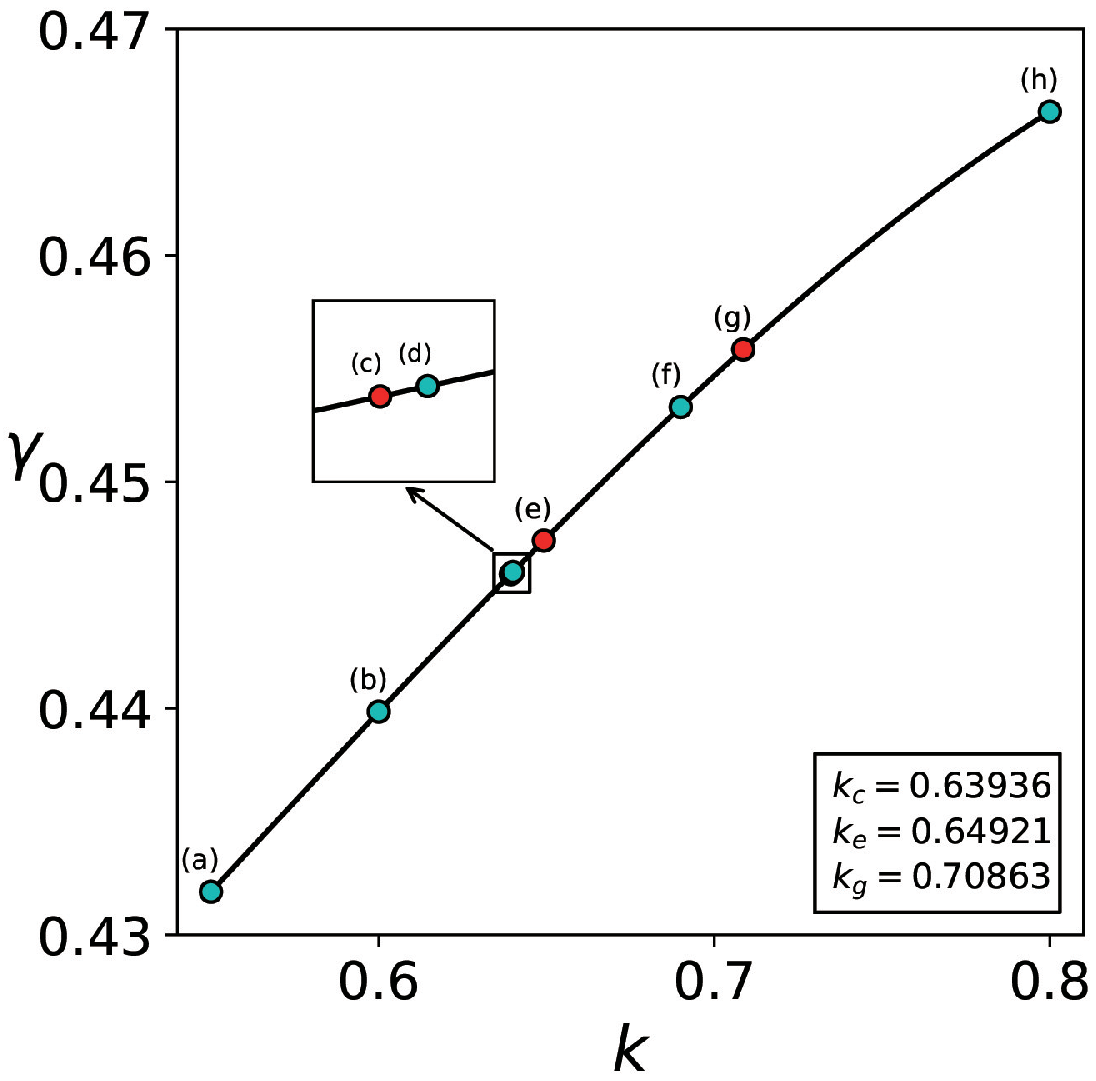}
    \includegraphics[width=0.397\textwidth]{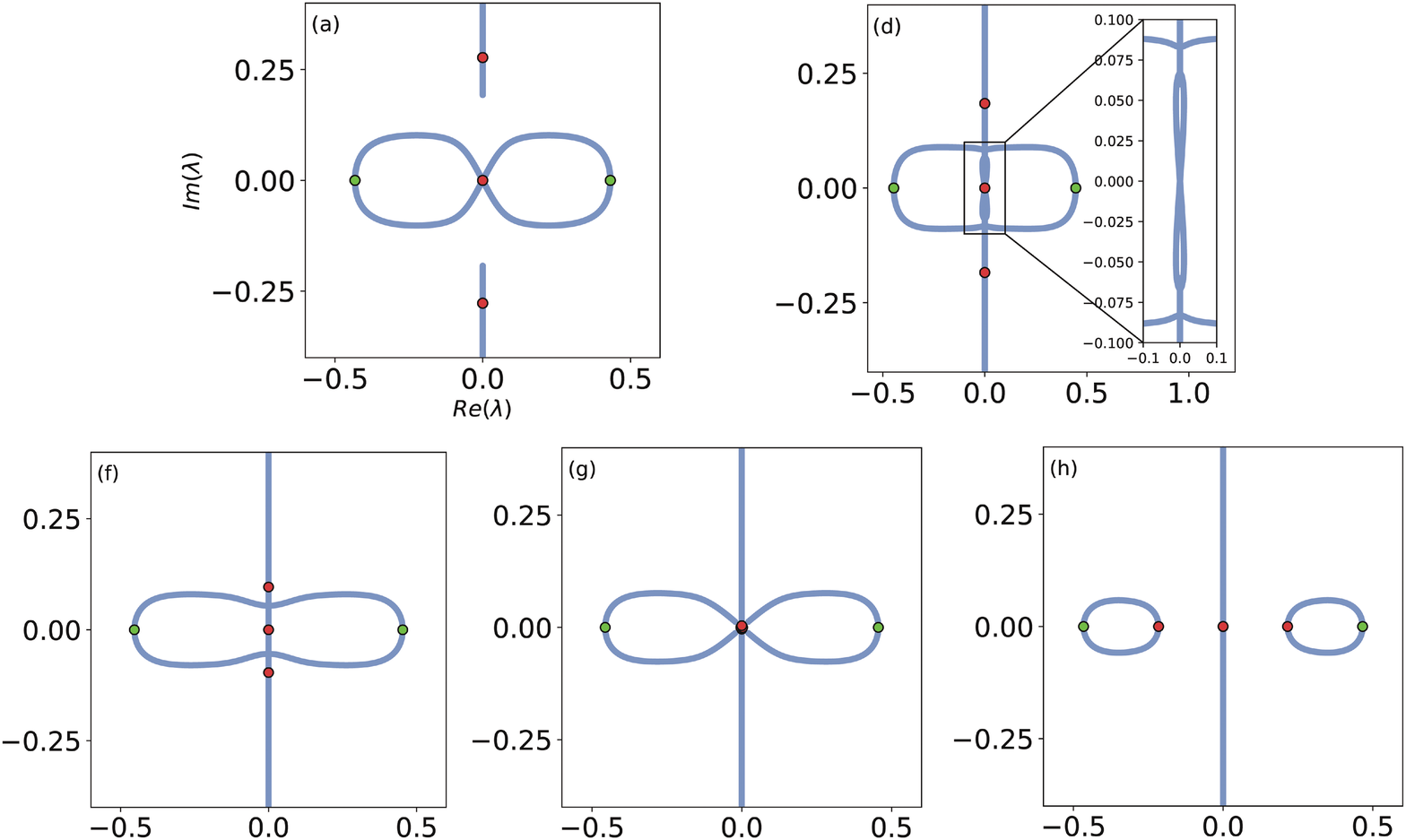}
    \vspace*{-0.05in}
    \caption{Left panel shows the maximal instability growth rate
      $\gamma$ as a function of $k$ with $B=0.25$ and $V_{0}=1$. When
      $k<k_{c}$, the modulation instability appears [one example can be seen in $(a)$ (with $k=0.55$) of the right panel]. At $k=k_{c}$, the ellipse-like curve appears. From $k_{c}$ to $k_{e}$, the dominant instability switches to the ellipse-like instability [one example can be seen in $(d)$ (with $k=0.64$) of the right panel]. At $k=k_{e}$,  MI disappears and only the ellipse-like eigenvalues exist. From $k_{e}$ to $k_{g}$, the  ellipse-like curve is compressed vertically [as shown in $(f)$ (with $k=0.69$) of the right panel]. An infinity symbol forms at $k_{g}$ (as shown in $(g)$ of the right panel). When $k>k_{g}$, the infinity symbol splits into two isolas drifting away along the real axis [as shown in $(h)$ (with $k=0.8$) of the right panel]. For the right panel, the red dots correspond to $\mu=0$ and the  green dots correspond to $\mu=\frac{\pi}{2K(k)}$.}
    \vspace*{-0.1in}
    \label{growth}
\end{figure*}

\begin{figure}[tpb]
    \centering
    \includegraphics[width=0.455\textwidth]{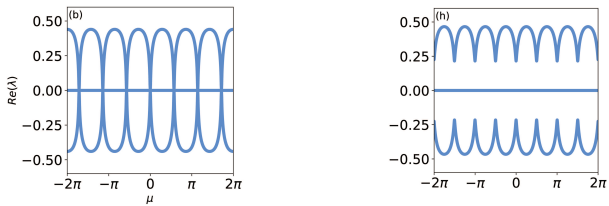}
    \vspace*{-0.1in}
    \caption{Real part of growth rates as a function of the Floquet
      parameter $\mu$, where $(b)$ (MI with $k=0.6$) and $(h)$ (isola
      instability with $k=0.8$) similarly to the left panel of FIG.~\ref{growth}}
    \vspace*{-0.1in}
    \label{growth1}
\end{figure}

\begin{figure}[tpb]
    \centering
    \includegraphics[width=0.455\textwidth]{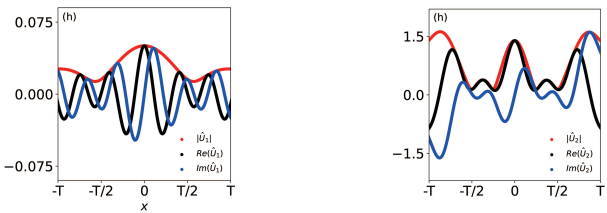}
    \vspace*{-0.1in}
    \caption{Eigenfunctions of the isola instability branch for $\mu=\frac{\pi}{2K(k)}$, where $(h)$ is labeled in FIG.~\ref{growth}.}
    \vspace*{-0.1in}
    \label{growth2}
\end{figure}

\subsection{\label{local} From MI to isola instability}

Note that the solutions~(\ref{solution1}) have three free parameters
$V_{0}$, $B$ and $k$. FIG.~\ref{fs1} shows the maximal instability
growth rate $\gamma$ as a function of $k$ and $V_{0}$ with
$B=0.25$. We can see that $\gamma$ increases with $k$ and $V_{0}$
increasing. To study the transition from MI to isola instability, by
fixing $V_{0}=1$ and $B=0.25$, we study the dynamics of instabilities
with varying $k$, i.e., effectively varying the periodicity of the potential.
As shown in FIG.~\ref{growth}, for  $0<k<k_{c}=0.63936$, the only
instability of the elliptic wave is the MI and the maximal instability
growth rate $\gamma$ corresponds to the real eigenvalues with
$\mu=\frac{\pi}{2K(k)}$, which implies that the dominant disturbance
has twice the period of elliptic wave. A typical example of MI is
shown in FIG.~\ref{growth}(a) with $k=0.55\in(0,k_{c})$ and it can be
seen that the closure of spectrum that
is not on the imaginary axis forms an infinity symbol centered at the
spectral plane origin. At $k=k_{c}$, the collisions of eigenvalues on the
imaginary axis (at $\pm 0.06202i$), lead to the appearance of an
ellipse-like curve. Then from $k_{c}$ to $k_{e}=0.64921$, two types of
instability appear, as shown in FIG.~\ref{growth}(d) with $k=0.64$,
which shows that a figure 8 is present inside an ellipse-like
curve. Therefore from $k_{c}$ to $k_{e}$, the dominant instability
switches to the ellipse-like instability, which is not
MI (recall that MI involves an unstable band of eigenvalues
encompassing the origin), and the dominant disturbance has twice
period of elliptic
waves. At $k=k_{e}$, a collision of eigenvalues at the origin leads to
the disappearance of MI and only the ellipse-like eigenvalues
exist. From $k_{e}$ to $k_{g}=0.70863$, the ellipse-like curve is
compressed vertically (see FIG.~\ref{growth}(f) with $k=0.69$).
Finally, this leads to the formation of an infinity symbol at
$k_{g}=0.70863$ (see FIG.~\ref{growth}(g)), again of the MI type. 

Increasing $k>k_{g}$,  the collision of the eigenvalues with
$\mu=0$ (red dots in FIG.~\ref{growth} (f,g,h)) at the origin (where a
Hamiltonian Hopf bifurcation occurs) causes the infinity symbol to
subsequently split into two isolas drifting away along the real axis,
as shown in Figure~\ref{growth}(h) with $k=0.8$. Now, the dominant
instability switches to the  isola instability, which is not MI (since
the latter involves the spectral plane origin),
and the dominant disturbance retains twice the period of the elliptic wave.

For the isola instability, we can observe the following.
(a) Similarly to MI,  the entire range of the Floquet parameter $\mu$
covers the isola instability branch (as shown in
FIG.~\ref{growth1}), in contrast to the high-frequency instabilities corresponding to a
narrow region of the Floquet parameter $\mu$ (as shown in
FIG.~\ref{growth4} below);
(b) Differently from the MI,  the spectrum of the isola instability
has no intersections with
the origin and the range of growth rate does not start from zero but
from the nonzero eigenvalues with $\mu=0$, as shown in
FIG.~\ref{growth1}; (c) The maximal instability growth rate
corresponds to the real eigenvalues with $\mu=\frac{\pi}{2K(k)}$ (as
shown in the green dots of FIG.~\ref{growth}(h)); (d) Such an isola
instability branch is called local instability branch in
fluids~\cite{ber1}, since the eigenfunctions associated with such a
branch change rapidly in the vicinity of the wave-crest. However, here
the eigenfunctions associated with such
a branch don't have such local property (which is a fundamental
difference in comparison to~\cite{ber1}), as shown in
FIG.~\ref{growth2}. Therefore, such
isola instability in BECs can be deemed to be nontrivially distinct
from the local instability branch in fluids.

\begin{figure}[tpb]
    \centering
    \includegraphics[width=0.25\textwidth]{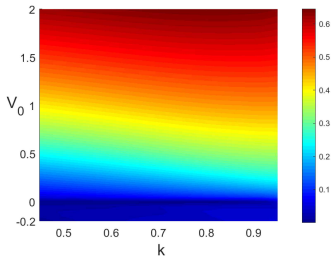}
    \vspace*{-0.1in}
    \caption{The maximal instability growth rate $\gamma$ as a function of $k$ and $V_{0}$ with $B=0.3$.}
    \vspace*{-0.1in}
    \label{fs2}
\end{figure}

\begin{figure*}
    \centering
    \includegraphics[width=0.245\textwidth]{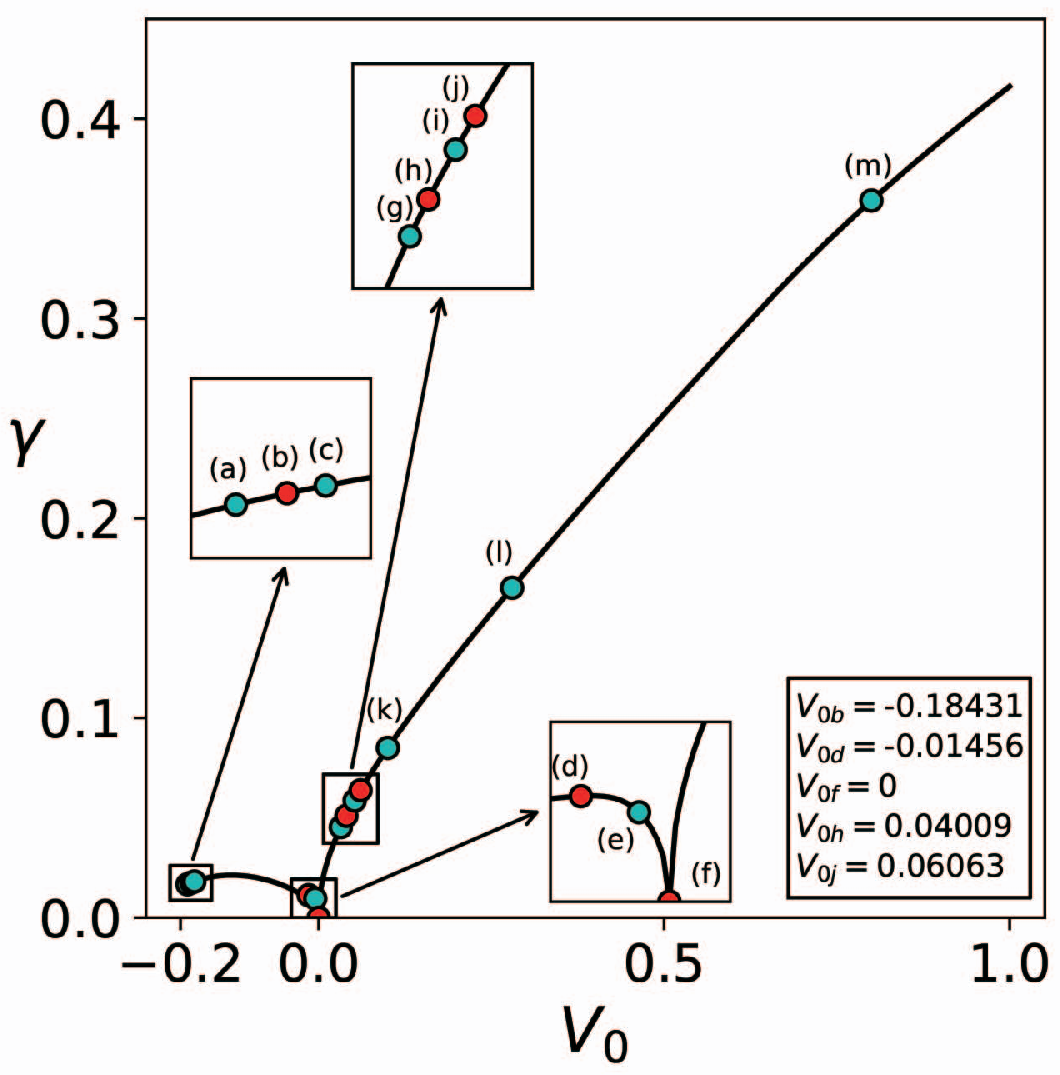}
    \includegraphics[width=0.497\textwidth]{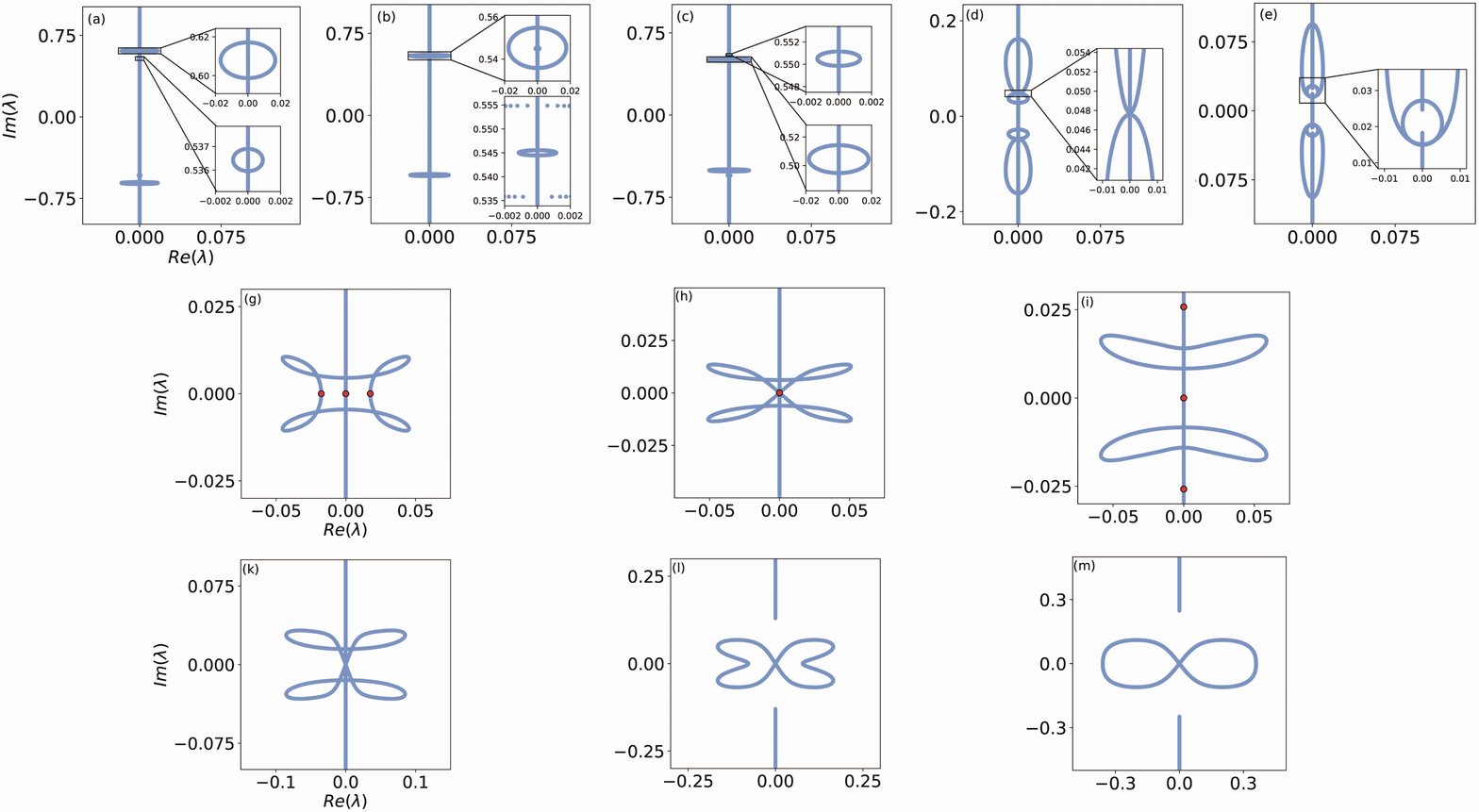}
    \vspace*{-0.05in}
    \caption{(Left panel) The maximal instability growth rate $\gamma$ as a function of $V_{0}$. When $V_{0}=V_{0f}$, the elliptic solutions are stable.
When $-0.19 <V_{0}<V_{0b}=-0.18431$, the first bubble dominates the
instability (see $(a)$ (with $V_{0}=-0.19$) of the right panel). From
$V_{0}$ to $V_{0b}$, the two bubbles approach and collision (see $(b)$
of the right panel). When $V_{0}>V_{0b}$, the two bubbles pass through
each other (see $(c)$ (with $V_{0}=-0.18$) of the right panel) and
then they fuse together (see $(d)$ (with $V_{0}=-0.01456$) of the right panel) and move toward the origin (see $(e)$ (with
$V_{0}=-0.005$) of the right panel).
When $0<V_{0}<V_{0j}=0.06063$, a transition between different stability spectra caused by the collision of eigenvalues with $\mu=0$ at the origin happens (see $(g,h,i)$ (with $V_{0}=0.032, 0.04009, 0.052$) of the right panel). The red dots in $(g,h,i)$  of the right panel correspond to $\mu=0$.  When $V_{0}>V_{0j}$, we show three different stability spectra (see $(k,l,m)$  (with $V_{0}=0.1, 0.28, 0.8$) of the right panel). 
    }
    \vspace*{-0.1in}
    \label{growth3}
\end{figure*}

\begin{figure}[tpb]
    \centering
    \includegraphics[width=0.25\textwidth]{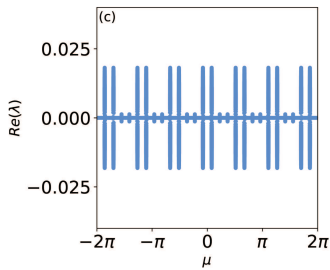}
    \vspace*{-0.1in}
    \caption{Real part of growth rates as a function of the Floquet parameter $\mu$, where $(c)$  (high-frequency instability) is labeled in the left panel of FIG.~\ref{growth3} }
    \vspace*{-0.1in}
    \label{growth4}
\end{figure}

\begin{figure}[tpb]
    \centering
   \includegraphics[width=0.25\textwidth]{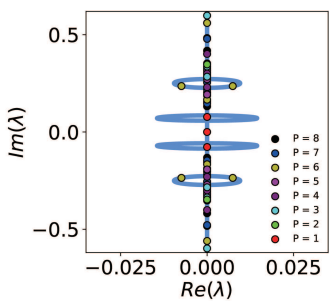}    \vspace*{-0.05in}
    \caption{The stability spectrum for the elliptic solutions~(\ref{solution1}) with respect to $P$-subharmonic perturbations ($k=0.5$, $B=0.3$ and $V_{0}=-0.04$).}
    \vspace*{-0.1in}
    \label{sta1}
\end{figure}

\begin{figure}[tpb]
    \centering
    \includegraphics[width=0.25\textwidth]{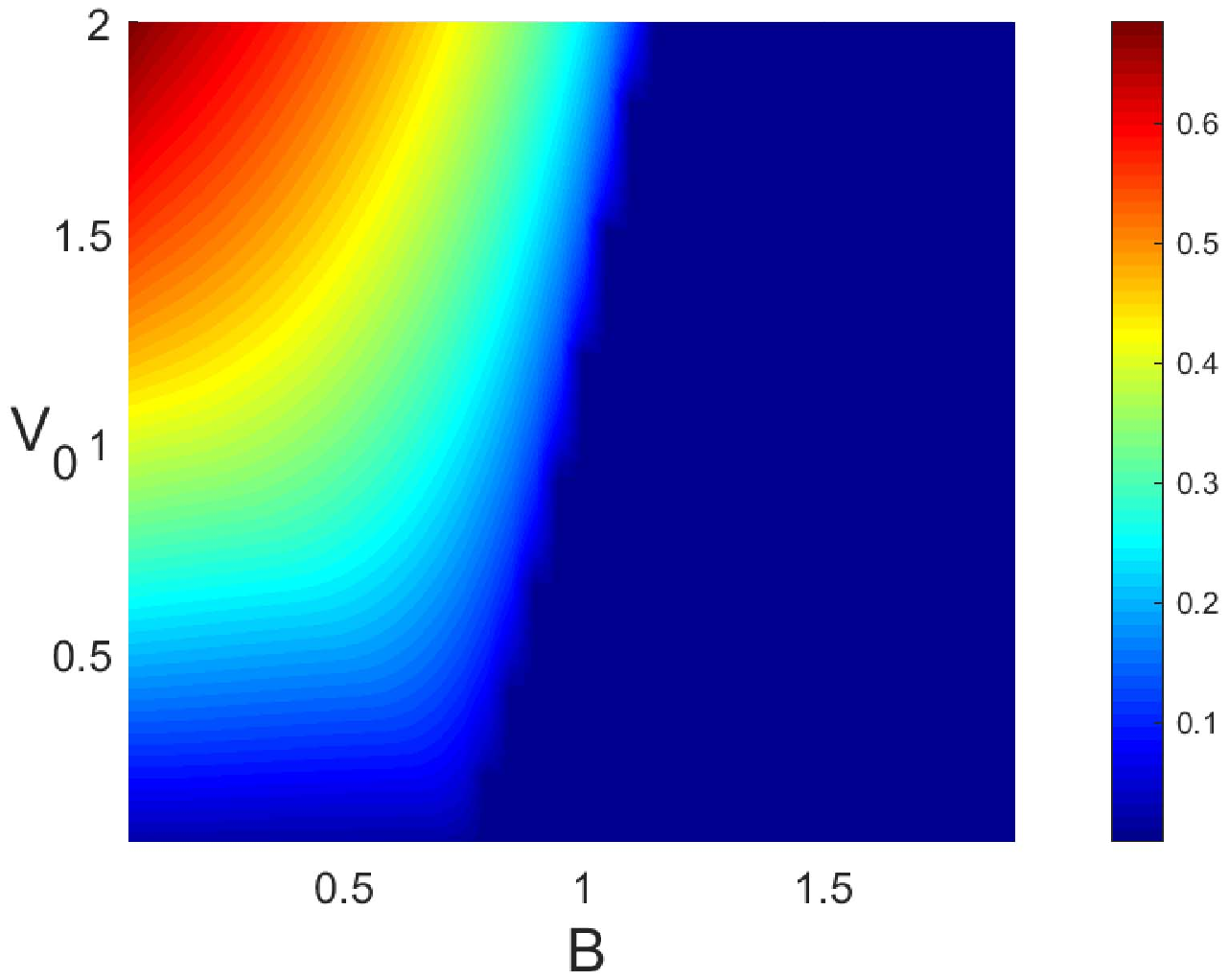}
    \vspace*{-0.1in}
    \caption{The maximal instability growth rate $\gamma$ as a function of $B$ and $V_{0}$ with $k=0$.}
    \vspace*{-0.1in}
    \label{fs3}
\end{figure}

\begin{figure*}
    \centering
        \includegraphics[width=0.195\textwidth]{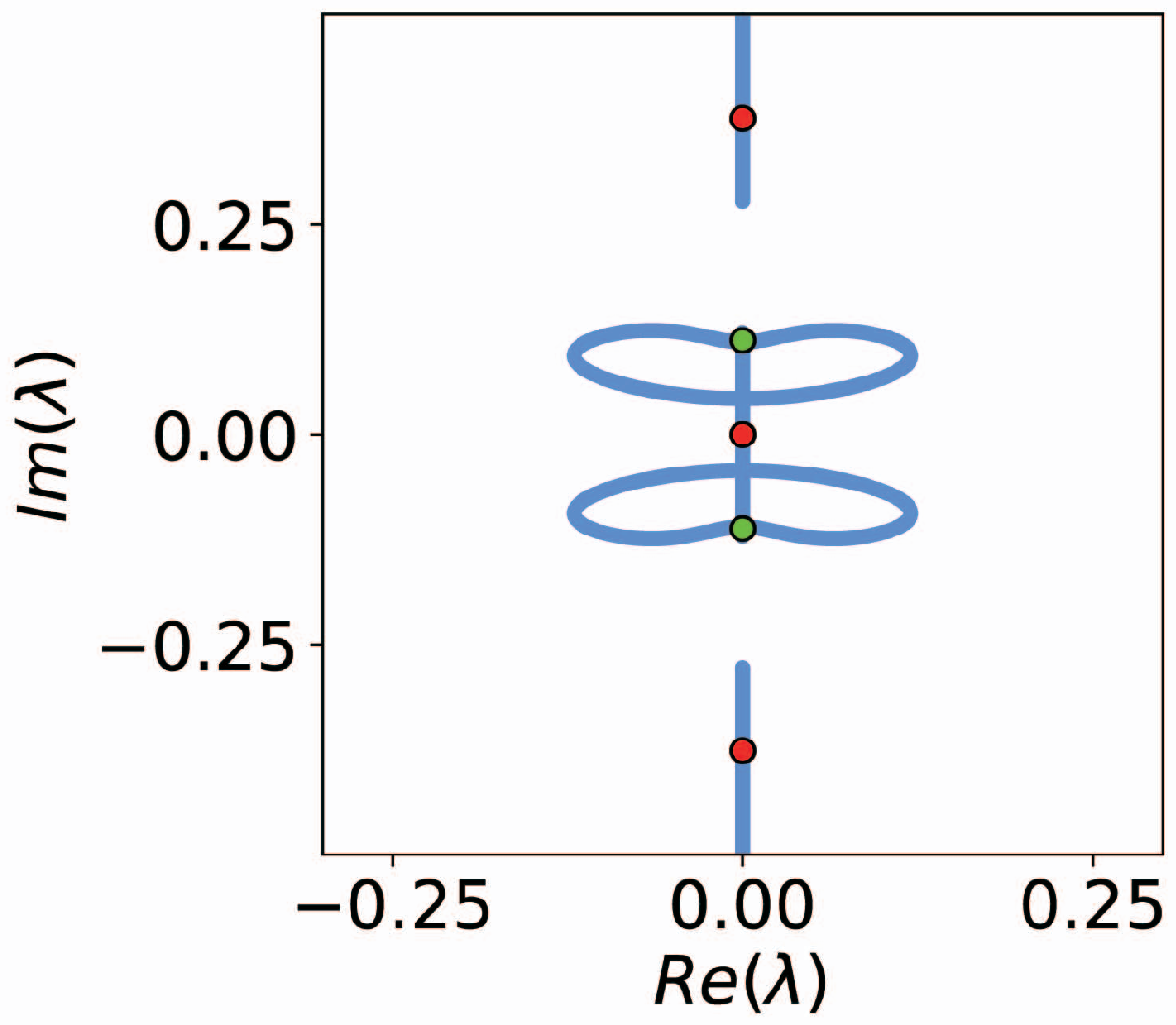}
    \includegraphics[width=0.19\textwidth]{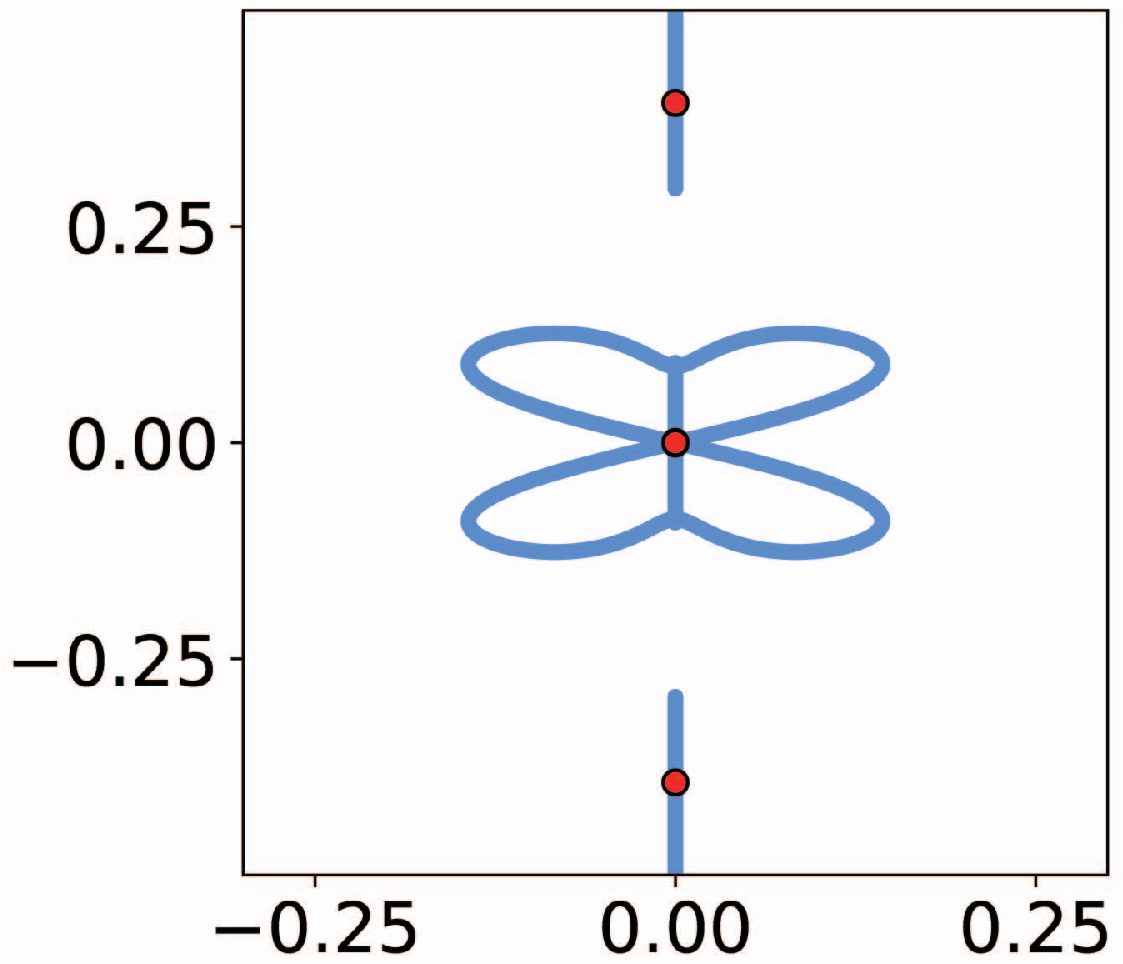}
     \includegraphics[width=0.185\textwidth]{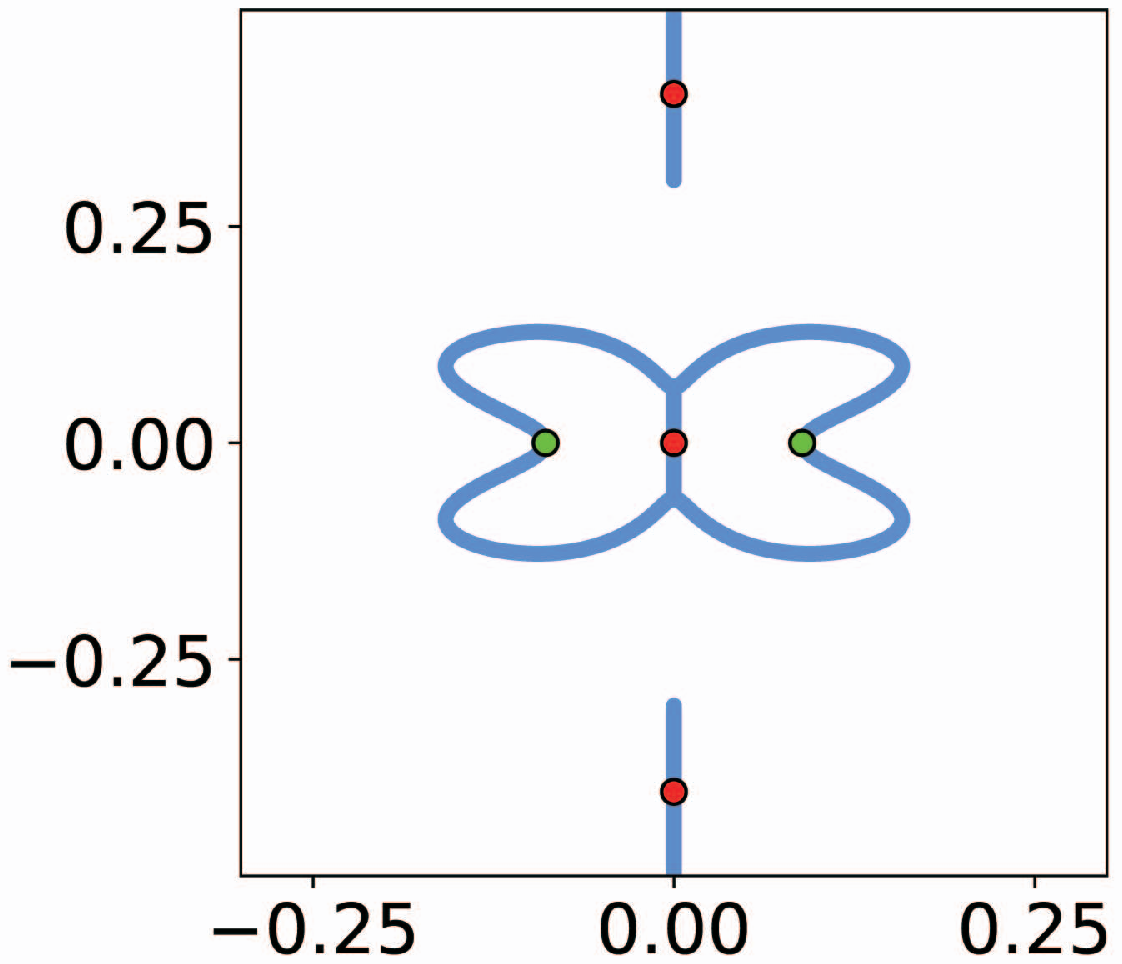}     
       \includegraphics[width=0.19\textwidth]{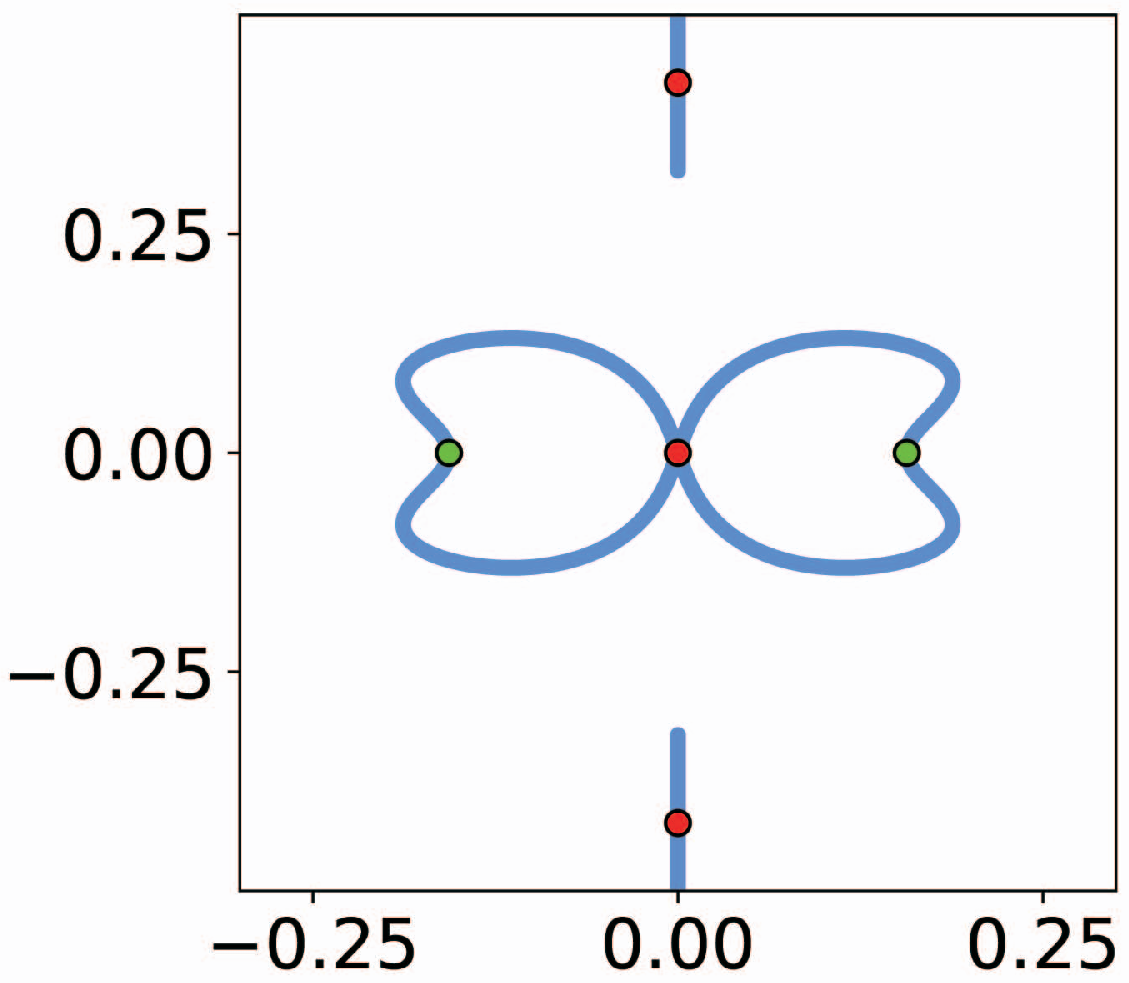}
     \includegraphics[width=0.18\textwidth]{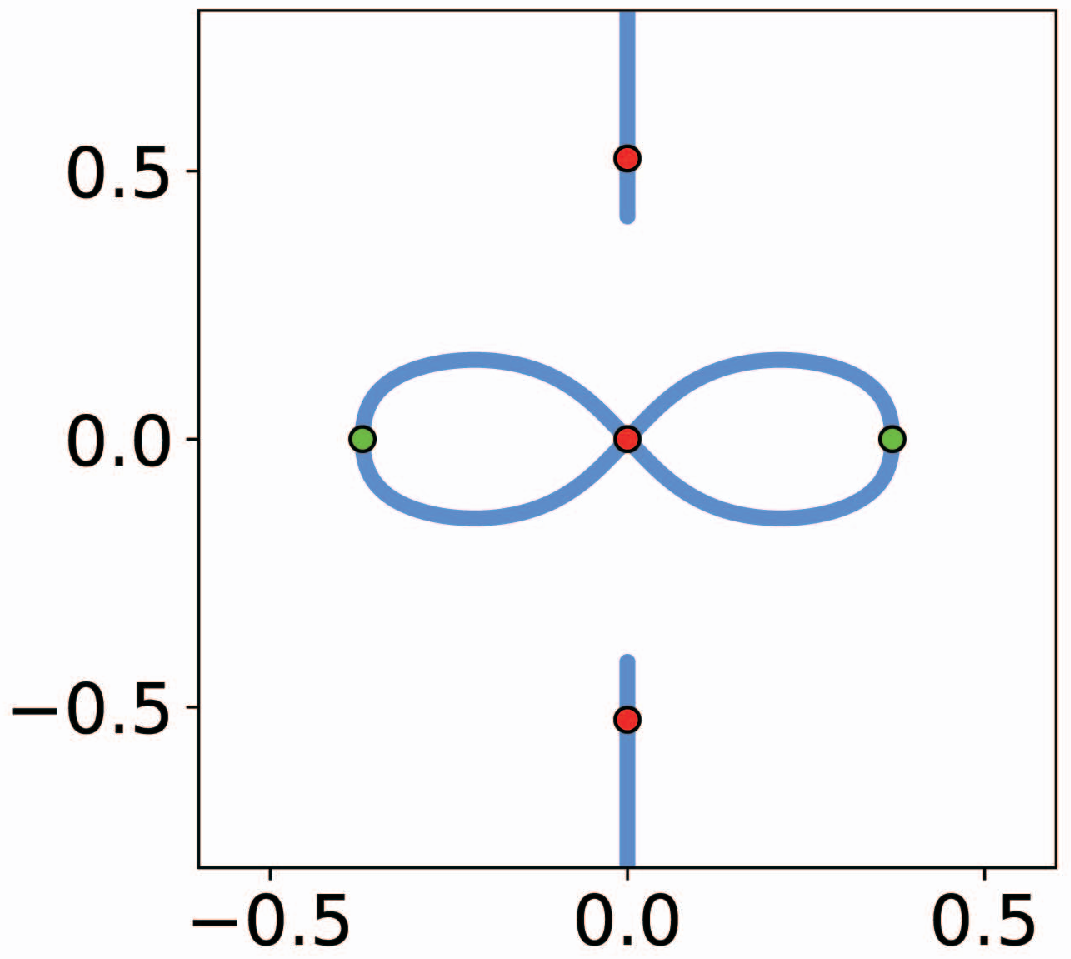}     
    \vspace*{-0.05in}
    \caption{The stability spectrum for the elliptic solutions~(\ref{solution1}) with $k=0$, $B=0.29$, (from left to right) $V_{0}=0.3$, $V_{0}=0.36166$, $V_{0}=0.4$, $V_{0}=0.48$ and $V_{0}=1$.}
    \vspace*{-0.1in}
    \label{growth9}
\end{figure*}

\subsection{\label{local} From MI to high-frequency instability}

FIG.~\ref{fs2} shows the maximal instability growth rate $\gamma$ as a function of $k$ and $V_{0}$ with $B=0.3$. We can see that $\gamma$ increases with $k$ and $V_{0}$ increasing. To study the transition from MI to high-frequency instability, by fixing $k=0.5$ and $B=0.3$, we study the dynamics of instabilities with varying $V_{0}$.
We note that when $V_{0}=V_{0f}=0$, (\ref{bec1}), the problem reduces
to the standard defocusing NLS equation. It is well known that all
elliptic solutions with $V_{0}=0$ are stable~\cite{adbe1}.
Here we consider the case where $V_{0}\neq 0$.
As shown in FIG.~\ref{growth3}, when $V_{0}<0$, only the
high-frequency instability occurs (see
FIG.~~\ref{growth3}(a,b,c,e)). The high-frequency instability (the
corresponding perturbations oscillate in time) develops  from a
Hamiltonian-Hopf bifurcation: collisions of nonzero, imaginary
elements of the stability spectrum ($V_{0}=0$) lead to  eigenvalues
symmetrically bifurcating from the imaginary axis as $V_{0}$
decreasing, resulting in instability.
Specifically, from $-0.19 \leqslant V_{0}<V_{0b}=-0.18431$, the first
bubble (arising
farther away from the origin) dominates the instability (see
FIG.~\ref{growth3}(a) with $V_{0}=-0.19$). By increasing $V_{0}$ to
$V_{0b}$, the two bubbles approach and collide (see
FIG.~\ref{growth3}(b)). When $V_{0}>V_{0b}$, the two bubbles pass
through each other (see FIG.~\ref{growth3}(c) with $V_{0}=-0.18$) and
subsequently they
fuse together (see FIG.~\ref{growth3}(d)
with $V_{0}=-0.01456$) and move toward the origin  (see FIG.~\ref{growth3}(e)
with $V_{0}=-0.005$).
When $0<V_{0}<V_{0j}=0.06063$, the elliptic solutions are modulational
stable. We can see the transition between different stability spectra
caused by the collision of eigenvalues with $\mu=0$ at the origin (see
FIG.~\ref{growth3}(g,h,i)). When $V_{0}>V_{0j}$, the modulation
instability appears and we show three different stability spectra (see
FIG.~\ref{growth3}(k,l,m)). Different from MI and isola instability,
we can see that the high-frequency instability branch occurs in a
narrow region of the Floquet parameter $\mu$ (as shown in
FIG.~\ref{growth4}), which implies that we may get some
useful stability results with respect to subharmonic
perturbations. For example, we show that the elliptic
solutions~(\ref{solution1}) (with $k=0.5$, $B=0.3$ and $V_{0}=-0.04$)
are stable with respect to $1$-, $2$-, $3$-, $4$-~, $5$-,  $7$- and
$8$- subharmonic perturbations but unstable with respect to the $6$- subharmonic perturbation, as shown FIG.~\ref{sta1}.

\subsection{\label{local} Instability trapped in a standing light wave $(k=0)$}

When $k=0$, the elliptic potential $V(x)$ reduces to the trigonometric functions and thus $V(x)$ is a standing light wave. 
FIG.~\ref{fs3} shows the maximal instability growth rate $\gamma$ as a function of $B$ and $V_{0}$ with $k=0$. We can see that $\gamma$ increases with $V_{0}$ increasing and $B$ decreasing.
The MI arises from the collision of the eigenvalues with
$\mu=\frac{\pi}{2K(k)}$ at the origin, as shown in~FIG.~\ref{growth9}.
Here, we also note that in~FIG.~\ref{growth9} a panel with four petals
morphs into  a panel with two
petals. This is because increasing $V_{0}$ leads to more collisions of
imaginary eigenvalues at the origin and the spectral curve is
 vertically compressed.
Besides we note that, solutions~(\ref{solution1}) with $k=0$ and
$b=0.29$ are stable with respect to co-periodic perturbations. When
$V_{0}$ is large enough, it can be seen that
the maximal instability growth rate corresponds to the real
eigenvalues with $\mu=\frac{\pi}{2K(k)}$, as shown in FIG.~\ref{growth9} with $V_{0}=1$.

\begin{figure}[tpb]
\centering
\includegraphics[height=85pt,width=80pt]{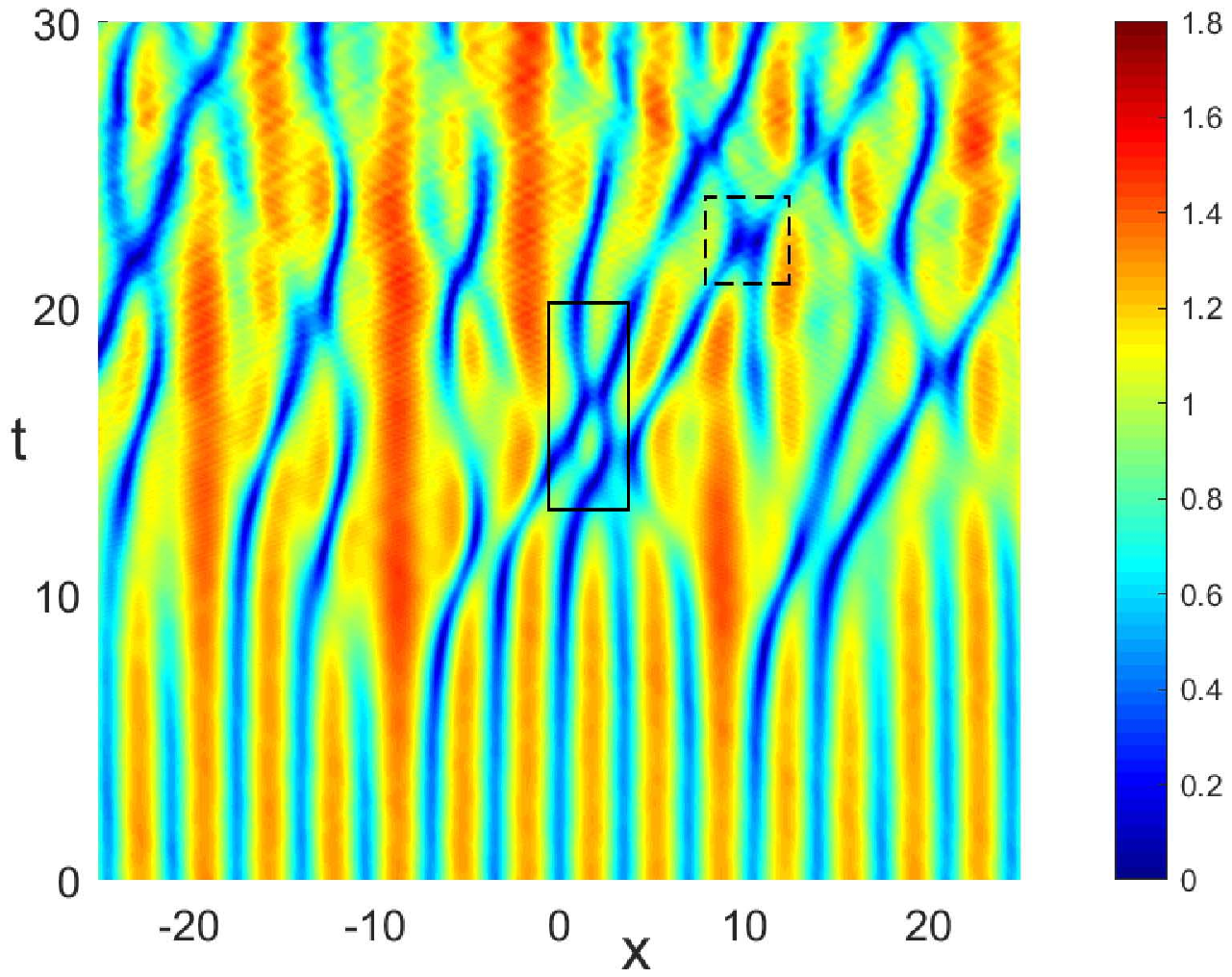}
\includegraphics[height=85pt,width=80pt]{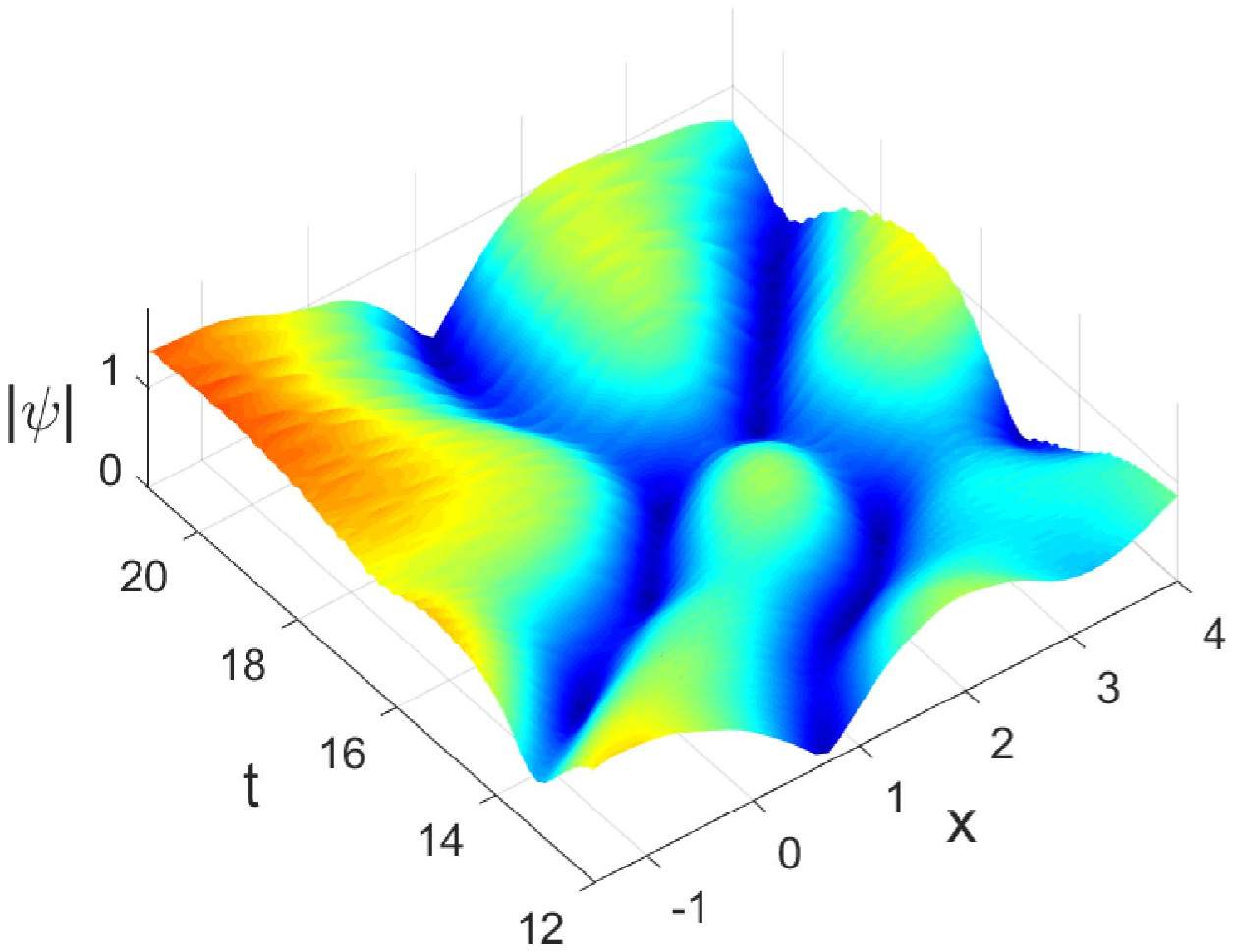}
\includegraphics[height=85pt,width=80pt]{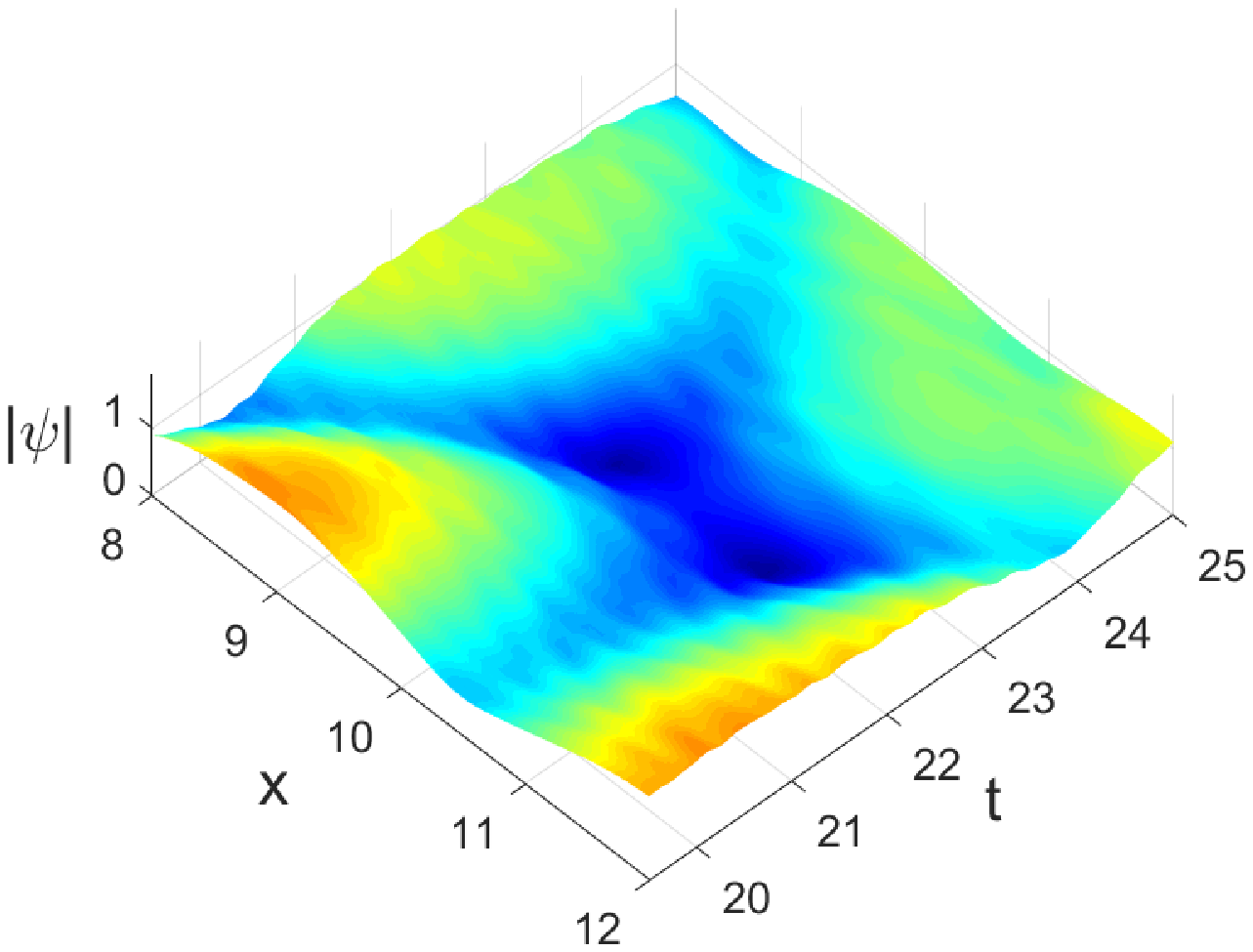}
\caption{\label{rw11}
  Numerical evolution of the MI resulting in pattern distortion and
  collisional
  events (highlighted by boxes on the left panel and zoomed-in at the
  middle
  and right panels). The initial condition is solution~(\ref{solution1}) perturbed by $5 \%$
random noise with $k=0.6$, $B=0.25$ and $V_{0}=1$. The amplitude evolution (upper-left);
the (solid-line)zoomed-in evolution of the box on the upper-left is shown on
the middle panel; the (dotted-line) zoomed-in evolution of the box on the left is shown on
the rightmost  panel.}
\end{figure}

\begin{figure}[tpb]
\centering
\includegraphics[height=85pt,width=90pt]{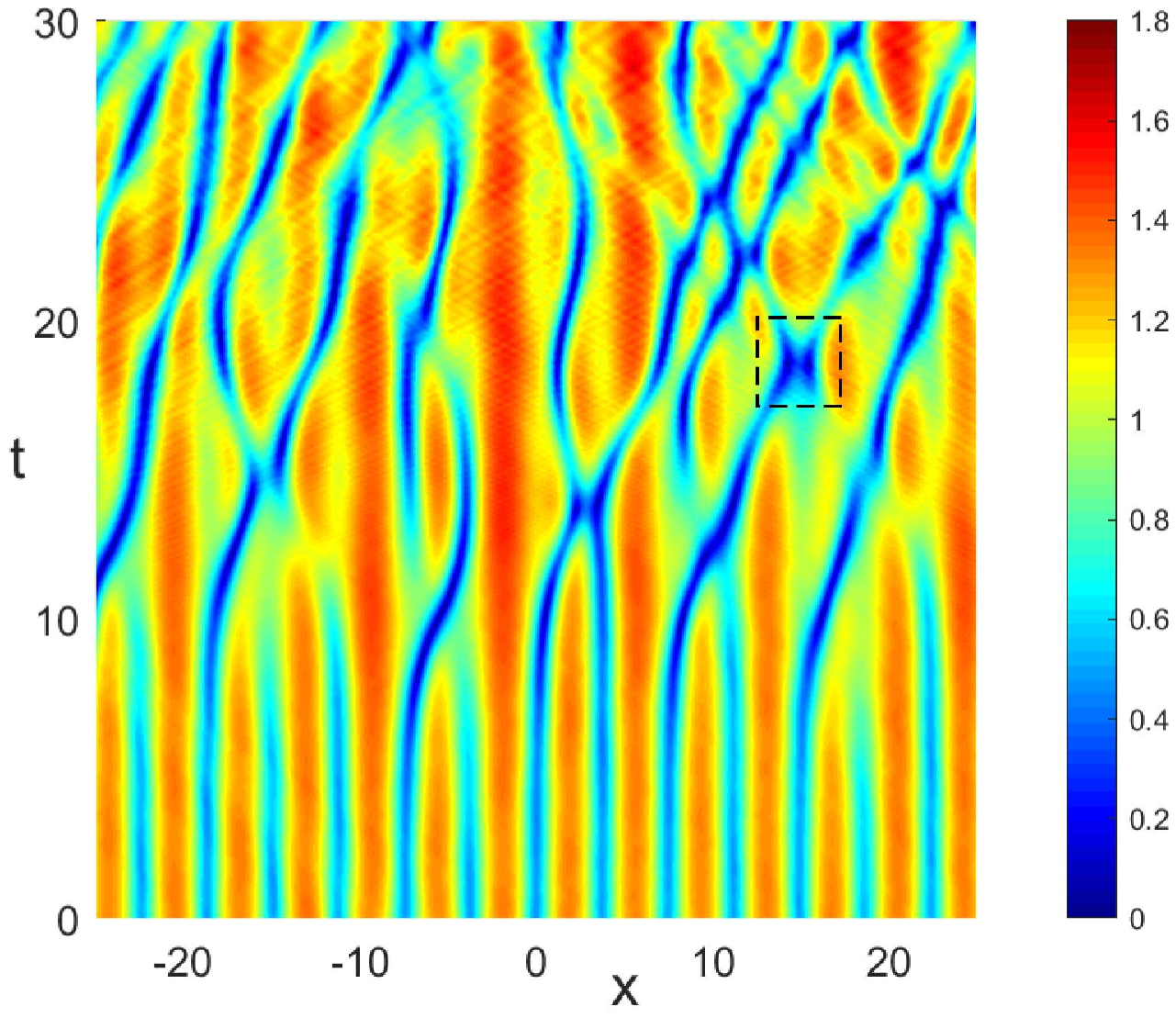}\quad \includegraphics[height=85pt,width=90pt]{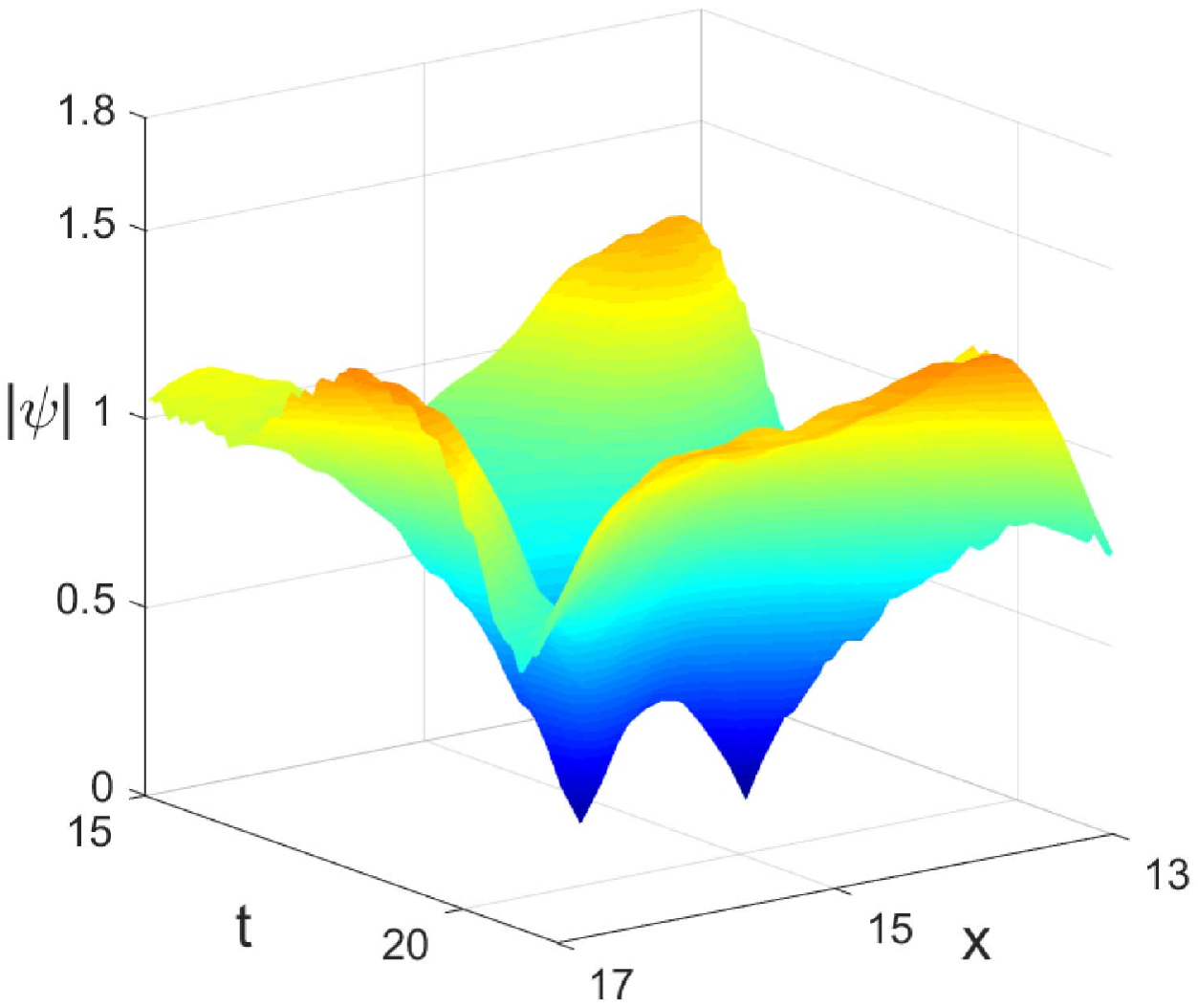}\\
\caption{\label{rw12}
Numerical excitation of the isola instability. The initial condition is solution~(\ref{solution1}) perturbed by $5 \%$
random noise with $k=0.72$, $B=0.25$ and $V_{0}=1$. The amplitude evolution (left);
the zoomed-in evolution of the box on the left is shown on
the right panel.}
\end{figure}

\subsection{\label{LS} Dynamical Manifestation of the Instabilities}

Having explored the different scenarios of instability, we now turn to
direct
numerical simulations in order to explore the dynamical byproducts of
these instabilities.
Starting from the nontrivial phase elliptic
solutions~(\ref{solution1}), we impose random perturbations and
visualize the patterns produced by~(\ref{bec1}) numerically. 


The case of dynamical evolution of a modulationally unstable
scenario is shown in Fig.~\ref{rw11}. One can observe that after an
initial stage, the periodic pattern is distorted leading to the
emergence
of some skewed density dips reminiscent of moving dark (gray)
solitary waves; for details of such coherent structures, see the
review of~\cite{djf}. As these structures move through the
distorted pattern they appear to interact in collision-type events
which are somewhat reminiscent of the interactions of dark solitary
waves observed, e.g., in the experiments
of~\cite{markus1,sengstock,markus2}.
These types of events causing a (deeper) spatio-temporal dip, prior to
the colliding patterns re-emerging are highlighted in two boxes in
Fig.~\ref{rw11}
whose evolution is presented in more detail in the additional panels
of the figure.

Interestingly, and as perhaps may be expected by the similar nature of
the
relevant instability (although the isola instability is detached from
the spectral plane origin), the dynamics of the isola instability is
similar
to that of MI. Indeed, the relevant dynamical manifestations can be
seen in Fig.~\ref{rw12}. Here, too, it is evident that
the distortion of the pattern leads to a number of waves that
propagate along skewed lines in the space-time (left) panel
of the figure. The zoom-in to the box of the left panel
is once again shown in the right panel, illustrating a space-time
collisional type event before the participating waves once again
separate.
It is relevant to note here that similar results to those of
Figs.~\ref{rw11}-\ref{rw12}
arise in the case of $k=0$, i.e., for a trigonometric standing wave of
light (results not shown here).

\begin{figure}[tpb]
\centering
\hspace{-1cm}\includegraphics[height=85pt,width=120pt]{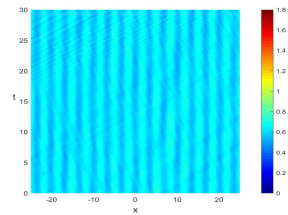}\\
\caption{\label{rw13}
The initial condition is solution~(\ref{solution1}) perturbed by $5 \%$
random noise with $k=0.5$, $B=0.3$ and $V_{0}=-0.1$.}
\end{figure}

Finally, it is interesting to point out the  fundamental difference of
the high-frequency instability,
in comparison, e.g., with those of Figs.~\ref{rw11}-\ref{rw12} above.
A prototypical example of the high-frequency instability is
shown in Fig.~\ref{rw13}.  We can see that the instability is
manifested through the propagation of high-wavenumber unstable
modes which, in turn, are weakly perturbing the (deeper) density dips
of the original configuration.
Nevertheless, the latter persist, avoiding the more intense
collisional events described above.

\section{Conclusions \& Future Challenges}

The instabilities of the nontrivial phase elliptic solutions in a
repulsive Bose-Einstein
condensate (BEC) with a periodic potential have been studied. Based on
the
defocusing nonlinear Schr\"{o}dinger (NLS) equation with an elliptic
function potential based on a sequence of fundamental earlier
works~\cite{bec11,bec12,bec13,bec14}, the MI, the similar to it isola
instability (on the real line)
and the rather different high-frequency instability 
have been observed numerically and have been elucidated
quantitatively. With varying parameters in solutions and equation,
instability transitions occur, e.g., through a Hamiltonian Hopf
bifurcation. Specifically, (i) increasing $k$, we have observed that
the MI switches to the isola instability and the dominant disturbances
have twice elliptic wave's period, corresponding to a Floquet exponent
$\mu=\frac{\pi}{2K(k)}$. The isola instability arises from the
collision of spectral elements at the origin with $\mu=0$; (ii) with
varying $V_{0}$, the transition between the MI and high-frequency
instability occurs. Different from the MI and isola instability where
the collisions of spectral elements happen at the origin, the
high-frequency instability
arises from pairwise collisions of nonzero, imaginary elements of the
stability spectrum; (iii) in the limit of sinusoidal potential, with
varying $V_{0}$, we have shown the MI occurs from a collision of
eigenvalues with $\mu=\frac{\pi}{2K(k)}$ at the origin; (iv) the
dynamical evolution of the relevant instabilities has been elucidated,
notably
leading in the case of the MI and isola instabilities to distortion of
the patterns and events resembling the collision of dark (gray)
solitary waves.

Admittedly, since the emergence of these fundamental
works~\cite{bec11,bec12,bec13,bec14}, numerous developments have
arisen in higher-dimensional BECs~\cite{bec2-3}, in multi-component
condensates~\cite{ueda1,ueda2}, as well as in the context of
long-range interactions~\cite{lahaye}. Extending the present
considerations
to these progressively more and more accessible settings would
be a natural next step for future studies.

\noindent {\bf Acknowledgements} This work has been supported by the
Fundamental Research Funds of the Central Universities (No.
230201606500048) and the National
Natural Science Foundation of China under Grant No.12205029. The work of P.G.K is supported by
the US National Science Foundation under Grants No. PHY-2110030 and
DMS-2204702).

\end{document}